\documentclass[aps,prc,amsmath,amsfonts,amssymb,twocolumn,floatfix]{revtex4-1}
\usepackage{graphicx,bm,graphics}
\usepackage{color}

\begin{document}

\title{\vspace*{-0.2in}Using full configuration interaction quantum Monte Carlo in a seniority zero space to investigate the correlation energy equivalence of pair coupled cluster doubles and doubly occupied configuration interaction}
\author{James~J.~Shepherd}
\email{jamesjshepherd@gmail.com}
\altaffiliation{Present address:~Department of Chemistry, Massachusetts Institute of Technology, Cambridge, MA, 02139}
\affiliation{Department of Chemistry and Department of Physics and Astronomy, Rice University, Houston, TX 77005-1892}

\author{Thomas~M.~Henderson}
\affiliation{Department of Chemistry and Department of Physics and Astronomy, Rice University, Houston, TX 77005-1892}

\author{Gustavo~E.~Scuseria}
\affiliation{Department of Chemistry and Department of Physics and Astronomy, Rice University, Houston, TX 77005-1892}

\begin{abstract}
Over the past few years pair coupled cluster doubles (pCCD) has shown promise for the description of strong correlation.  This promise is related to its apparent ability to match results from doubly occupied configuration interaction (DOCI), even though the latter method has exponential computational cost.  Here, by modifying the full configuration interaction quantum Monte Carlo (FCIQMC) algorithm to sample only the seniority zero sector of Hilbert space, we show that the DOCI and pCCD energies are in agreement for a variety of 2D Hubbard models, including for systems well out of reach for conventional configuration interaction algorithms.  Our calculations are aided by the sign problem being much reduced in the seniority zero space compared with the full space.  We present evidence for this, and then discuss the sign problem in terms of the wave function of the system which appears to have a simplified sign structure.
\end{abstract}
\date{\today}
\maketitle

\section{Introduction}
Nowadays the accurate and reliable theoretical description of weakly correlated electronic systems has become more or less routine.  Unfortunately, some of the most interesting and technologically relevant aspects of electronic behavior arise due to \textit{strong} correlations, and the theoretical description of strongly correlated electronic systems is far more challenging.  Loosely, this difficulty arises because the mean-field picture which underlies many traditional wave function techniques is invalid in strongly correlated systems, so the conventional paradigm of describing a system in terms of particle-hole excitations out of a mean-field state breaks down.  The description of strongly correlated systems, in other words, requires an alternative way of organizing Hilbert space.

Recently, an increasing amount of evidence has suggested that the concept of seniority may be of significant use in this regard.  For our purposes here, the seniority of a determinant can just be defined as the number of singly-occupied spatial orbitals within that determinant.  In many cases, it appears that one can describe strongly correlated systems in terms of determinants with low seniority but perhaps high excitation rank with respect to one another (see, for example, Ref. \onlinecite{Bytautas2011}).

Conceptually, perhaps the simplest approach to including the determinants relevant for strong correlation is to use the doubly-occupied configuration interaction\cite{Allen1962,Smith1965,Veillard1967,Weinhold1967,Couty1997,Kollmar2003,Bytautas2011}  (DOCI), in which one simply diagonalizes the Hamiltonian in the basis of all seniority zero determinants.  This eliminates the need to determine a reference determinant or to pick an excitation level at which to truncate the wave function, and, with the right orbitals, often yields quite reasonable results.  It has, however, a crucial drawback: the computational cost of a DOCI calculation scales combinatorially with system size.  Worse yet, the results of a DOCI calculation are not invariant to rotations amongst the orbitals, so in practice some form of orbital optimization is required.  Together, these two limitations suggest that DOCI is, in practice, of extremely limited utility.

However, not all hope for DOCI-quality calculations is lost.  Somewhat unexpectedly, a method we will refer to as pair coupled cluster doubles\cite{Stein2014,Henderson2014b,henderson_pair_2015} (pCCD) and which was first called the antisymmetric product of 1-reference orbital geminals\cite{Limacher2013,Limacher2014,Tecmer2014,Boguslawski2014,Boguslawski2014a,Boguslawski2014b,Tecmer2015,Boguslawski2015} (AP1roG) provides results which nearly match those of DOCI, but with mean-field computational scaling.  To the extent that this affordable method matches DOCI at a fraction of the cost, and to the extent that DOCI accurately describes strongly correlated systems, pCCD shows considerable promise.

It is not our purpose here to assess the accuracy of pCCD or of DOCI.  Rather, we wish to check the first of the two criteria: that pCCD reproduces DOCI.  Thus far, this has been well established for small systems for which DOCI is computationally feasible.  Here, we wish to compare the two methods for much larger systems than DOCI has hitherto reached, to verify that the agreement between DOCI and pCCD is not simply an artefact of small system size. To do so, we will use a version of full configuration interaction quantum Monte Carlo (FCIQMC), modified to sample the seniority zero space and using the basis set from an optimized pCCD calculation. We also wish to demonstrate the utility of the seniority concept in FCIQMC calculations, in that the sign problem is substantially ameliorated in the seniority zero space.  We analyze the way in which this happens before leveraging this to address 2D Hubbard model systems. 

\subsection{Pair CC}
Put briefly, the idea in pCCD is to write the wave function as
\begin{equation}
|\mathrm{pCCD}\rangle = \mathrm{e}^T |0\rangle
\end{equation}
where $|0\rangle$ is a closed-shell single determinant and the cluster operator $T$ is
\begin{equation}
T = \sum_{ia} t_i^a \, c_{a_\uparrow}^\dagger \, c_{a_\downarrow}^\dagger \, c_{i_\downarrow} \, c_{i_\uparrow}
\end{equation}
with indices $i$ and $a$ respectively referring to spatial orbitals occupied and empty in $|0\rangle$.  Note that we have only $o v$ cluster amplitudes, where $o$ and $v$ are the number of occupied and virtual orbitals, respectively.

One can follow all the usual machinery of coupled cluster theory from here.  For example, one could introduce a left-hand wave function
\begin{equation}
\langle \overline{\mathrm{pCCD}} | = \langle 0 | (1 + Z) \, \mathrm{e}^{-T}
\end{equation}
where
\begin{equation}
Z = \sum_{ia} z^i_a \, c_{i_\uparrow}^\dagger \, c_{i_\downarrow}^\dagger \, c_{a_\downarrow} \, c_{a_\uparrow}
\end{equation}
and use this wave function to evaluate expectation values (including the energy).  Thus, in terms of these left- and right-hand states, one would have
\begin{equation}
E = \langle \overline{\mathrm{pCCD}} | H | \mathrm{pCCD}\rangle = \langle 0| \left(1 + Z\right) \, \mathrm{e}^{-T} \, H \, \mathrm{e}^T |0\rangle.
\end{equation}
The amplitudes defining $T$ and $Z$ can be found by making the energy stationary:
\begin{equation}
\frac{\partial E}{\partial t_i^a} = \frac{\partial E}{\partial z^i_a} = 0.
\end{equation}

In addition to defining the energy, the left- and right-hand wave functions in pCCD can be used to define one-particle and two-particle reduced density matrices, generically through
\begin{equation}
\Gamma = \langle 0 | (1 + Z) \, \mathrm{e}^{-T} \, \hat{\Gamma} \, \mathrm{e}^T | 0\rangle
\label{defGamma}
\end{equation}
where $\hat{\Gamma}$ is short-hand for a string of creation and annihilation operators.  These density matrices are sparse and have a simple structure.\cite{Henderson2014b}  They can be used for expectation values, but perhaps more importantly for finding the appropriate set of orbitals.  To accomplish this latter feat, one can introduce the antihermitian one-body operator
\begin{equation}
\kappa = \sum_{p > q} \sum_\sigma \kappa_{pq} \, \left(c_{p_\sigma}^\dagger \, c_{q_\sigma} - c_{q_\sigma}^\dagger \, c_{p_\sigma}\right)
\end{equation}
where $p$ and $q$ index arbitrary spatial orbitals and $\sigma$ indexes spins.  Exponentiating this operator creates a unitary orbital transformation.  One can generalize the energy to
\begin{equation}
E(\kappa) = \langle 0 | (1 + Z) \, \mathrm{e}^{-T} \, \mathrm{e}^{-\kappa} \, H \, \mathrm{e}^\kappa \, \mathrm{e}^T |0\rangle
\end{equation}
and make the energy stationary with respect to the amplitudes $\kappa_{pq}$ to define the orbitals; the resulting expression depends on the one- and two-electron integrals defining the Hamiltonian and the one- and two-body density matrices given schematically in Eqn. \ref{defGamma}.  Typically, the resulting orbitals are localized, particularly in the presence of strong correlation, and in this localized basis pCCD tends to adopt a structure not too dissimilar to perfect pairing generalized valence bond, in which for each occupied orbital $i$ there is one virtual orbital $a$ for which $t_i^a$ is large.

We have noted that, for the small systems for which DOCI is feasible, pCCD and DOCI give essentially the same results.  This appears to be a consequence of a close match between the right-hand state $|\mathrm{pCCD}\rangle = \exp(T) |0\rangle$ and the DOCI wave function $\mathrm{DOCI}\rangle$.\cite{Henderson2014b}  For strongly correlated systems, the left-hand state of pCCD less accurately reproduces the DOCI wave function.  This has implications for the calculation of properties within pCCD, which matches DOCI because $|\mathrm{pCCD}\rangle \approx |\mathrm{DOCI}\rangle$ and
\begin{equation}
H |\mathrm{DOCI}\rangle = E_\mathrm{DOCI} |\mathrm{DOCI}\rangle + |\Omega \neq 0\rangle,
\end{equation}
where $|\Omega \neq 0\rangle$ has non-zero seniority.  Accordingly, the projective energy
\begin{equation}
E_\mathrm{pCCD} = \langle 0 | H | \mathrm{pCCD}\rangle \approx \langle 0 | H | \mathrm{DOCI}\rangle
\end{equation}
approximately matches the DOCI energy, because $\langle0|\Omega~\neq 0\rangle=0$.  The poor agreement between the pCCD and DOCI left-hand states might be expected to compromise the agreement between pCCD and DOCI for properties other than the energy.  However, we do not test this here because extracting density matrices from FCIQMC, while possible,\cite{OVery2014} is far from straightforward.

\subsection{FCIQMC}
Full configuration interaction quantum Monte Carlo\cite{booth_fermion_2009} (FCIQMC) finds the ground state for a Hamiltonian written in a basis of orthogonal Slater determinants:
\begin{equation}
H_{ij}=\langle D_i | \hat{H} | D_j \rangle .
\end{equation}
It does so by using an imaginary time projector quantum Monte Carlo method,
\begin{equation}
| \Psi (\tau) \rangle = e^{-\tau \hat{H}} | D_0 \rangle ,
\end{equation}
where $D_0$ is a reference determinant. The exact ground state is recovered in the long $\tau$ limit, provided that $D_0$ has non-zero overlap with this state.  In FCIQMC, these equations are simulated as
\begin{equation}
    c_{\bm{i}}(\tau+\delta\tau) = c_{\bm{i}}(\tau) - \delta\tau \sum_{\bm{j}} ( H_{\bm{ij}} -E_S (\tau) \delta_{ij} ) c_{\bm{j}}(\tau),
            \label{eqn_fciqmc}
\end{equation}
wherein the coefficient vector relates to the wavefunction:
\begin{equation}
| \Psi \rangle =\sum_i c_i | D_i \rangle .
\end{equation}
The energy offset $E_S (\tau)$ is an energy offset applied uniformly to the diagonal of the matrix which conserves normalization. 

Certain algorithms are then adopted such that these equations can be sampled in such a way that the vector does not immediately expand to fill the whole of the space.\cite{booth_linear-scaling_2014,petruzielo_semistochastic_2012,booth_fermion_2009}  The coefficients themselves are simulated as discrete signed walkers within a simulation and they interact and propagate according to Eq. \ref{eqn_fciqmc}.

The off-site events, when $j\neq i$ in the sum, are termed spawning events. The sum over $j$ is sampled once per Monte Carlo iteration, over non-zero $H_{ij}$.  The change in site $j$ due to a walker at $i$ is then given by
\begin{equation}
p= \frac{-\delta\tau H_{ij}}{p(j|i)}
\end{equation}
where $p(j|i)$ is the conditional probability of choosing $j$ given $i$.  The sign of the sites is resolved by the negative sign in this equation and the sign of $H_{ij}$.  When all $H_{ij}$ are negative, all of the coefficients can be chosen such that they have the same sign.  In this event, FCIQMC does not have a sign problem, and convergence is rapid.\cite{spencer_sign_2012}  More comparison is made with other quantum Monte Carlo methods in a recent review.\cite{umrigar_observations_2015}

The sign problem is a general problem in linear algebra where parameters of a solution can vary in sign.  Here, the coefficients of the wavefunction typically all have different signs that need to be resolved.  The sites need to discover their sign in the ground state of the problem over the course of the simulation.  This is achieved by cancelling walkers when a site has more than one sign of walkers, a process termed annhilation.  Appropriate sampling of annhilation events is critical for FCIQMC and for overcoming the sign problem.\cite{spencer_sign_2012,kolodrubetz_partial_2012,kolodrubetz_effect_2013,shepherd_sign_2014}

Other on-site events relate to the energy of the determinant ($H_{ii}$) and the energy offset $E_S$, and can either cause the number of walkers on a site to rise or fall.  Initially, $E_S$ is set to some number greater than the ground state energy, such as the Hartree--Fock energy, and this causes the population to grow.  When the simulation has reached a certain point, this growth can be curtailed by varying $E_S$ to maintain the norm of the wavefunction.  When the equations are simulated in steady-state with a sufficient number of walkers to appropriately represent these equations, $E_S$ is an estimator (termed the `shift' estimator) for the ground state energy.

The other estimator for the energy can be found by projecting the instantaneous wavefunction against a reference (here, the Hartree--Fock determinant),
\begin{equation}
E_P=\frac{ \langle \Psi \left( \tau \right) | H | \Psi_\mathrm{ref} \rangle} {\langle \Psi \left( \tau \right) | \Psi_\mathrm{ref} \rangle} ,
\label{proje}
\end{equation}
and the two energy measures $E_S$ and $E_P$ can be expected to agree in an appropriately well-conditioned and converged calculation.

It is important to note that there is an adaption of the algorithm called the initiator approximation.\cite{cleland_communications:_2010}  This imposes certain restrictions on the sign of the spawning events and greatly improves convergence to the ground state.  The drawback is that the simulation must now be converged much more stringently with walker number, and analysis of this is performed elsewhere and beyond the scope of this paper.\cite{shepherd_sign_2014,petruzielo_semistochastic_2012,booth_breaking_2011,booth_explicitly_2012}  
In practice, the initiator approach enables systems of far larger sizes to be investigated,\cite{shepherd_full_2012,booth_towards_2013,thomas_accurate_2015,schwarz_insights_2015} further increased by a semi-stochastic adaptation.\cite{petruzielo_semistochastic_2012,blunt_semi-stochastic_2015}

\subsection{The DOCI Hamiltonian}

Since DOCI is an expansion only in terms of determinants with seniority zero, it only uses a portion of the Hamiltonian; in other words, we can think of DOCI as exactly solving what we will call the seniority-preserving Hamiltonian $H_{\delta\Omega=0}$, which is given by\cite{henderson_pair_2015}
\begin{equation}
H_{\delta\Omega=0}
 = \sum_p h_p \, N_p
 + \frac{1}{4} \, \sum_{p \neq q} w_{pq} \, N_p \, N_q
 + \sum_{pq} v_{pq} \, P_p^\dagger \, P_q.
\end{equation}
Here, the integrals are
\begin{subequations}
\begin{align}
h_p &= \langle p | h | p\rangle,
\\
v_{pq} &= \langle pp | v | qq \rangle,
\\
w_{pq} &= 2 \, \langle pq | v | pq \rangle - \langle pq | v | qp \rangle,
\end{align}
\end{subequations}
in terms of one-electron integrals $\langle p | h | q \rangle$ and two-electron integrals $\langle pq | v |rs \rangle$ expressed in Dirac notation.  The number and pairing operators $N_p$ and $P_p$ are given by
\begin{subequations}
\begin{align}
N_p &= \sum_\sigma a_{p_\sigma}^\dagger \, a_{p_\sigma},
\\
P_p &= a_{p_\downarrow} \, a_{p_\uparrow}.
\end{align}
\end{subequations}

We should emphasize that DOCI depends critically on the basis with respect to which seniority is defined.\cite{Limacher2014,Limacher2014,Stein2014,Henderson2014b,Boguslawski2014a,Boguslawski2014b}  

In the Hubbard Hamiltonian, if we were to define seniority with respect to the canonical plane wave basis, we would include all of the one-electron Hamiltonian because it is diagonal in this basis; if we were instead to define seniority with respect to the site basis, we would include none of the one-electron Hamiltonian because in this basis its diagonal blocks vanish.  In practice, the choice of basis should be optimized, and we have done so for the calculations here.  The resulting basis is essentially comprised of two-site (``dimer'') functions.

\section{DOCIQMC and the sign problem}
In order to find the doubly occupied configuration interaction (DOCI) solution, we set up FCIQMC to sample the seniority zero sector of the FCI space.  Orbital relaxation between pCCD and DOCI is not addressed, but should not be resolvable within the stochastic error of the quantum Monte Carlo simulations in any event.  Assuming prior optimization of the orbitals, this is a very straight forward modification from the point of FCIQMC.  Starting with a seniority zero reference, we can simply discard the matrix elements $H_{ij}$ that lead to a change in seniority.  This amounts to zeroing out the elements of the electron repulsion integrals to only allow values where the orbitals are in the pairing convention.  In other words, one retains only integrals of the form $\langle p p| v |q q\rangle$ for molecular orbitals $p$ and $q$.  Although it is possible to conceive of more sophisticated algorithms for doing this in general, for the purposes of this study we were able to reach the desired system size with this more simple approach.  We retain also integrals of the form $w_{pq}$, which in the DOCI case appear only in the diagonal entries of the Hamiltonian.

In general, the bottle-neck for this procedure would be orbital optimization on the fly within FCIQMC.
Orbital optimization within the context of this QMC method has recently been made possible, though its detailed discussion is beyond the scope of this manuscript~\cite{thomas_stochastic_2015}.  
We recall, however, that the  scaling of DOCIQMC is substantially worse than that of pCCD. 
So, as previous results make clear, essentially equivalent results can be found by instead optimizing orbitals via pCCD, for which the orbital optimization is not too expensive~\cite{henderson_pair_2015}.
While in principle orbitals optimized with DOCI and orbitals optimized with pCCD may differ, observationally these differences are very small and do not change the results of the deterministic DOCI calculation to more than a few microHartree.  As this difference due to orbital choice is well below the statistical error in converging our DOCIQMC calculations, we do not reoptimize orbitals at the DOCIQMC level.

\begin{figure}
\includegraphics[width=0.4\textwidth]{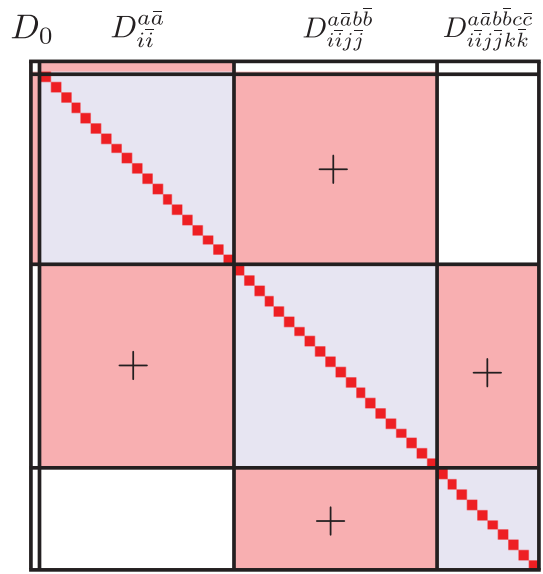}
\caption{A schematic representation of a partial seniority zero Hamiltonian, where only positive or zero matrix elements separate the off-diagonal blocks (marked $+$)  which connect different excitation ranks (e.g. doubles and quadruples). The rank of the matrix has been truncated for brevity, only containing doubles ($D_{i\bar{i}}^{a\bar{a}}$), quadruples ($D_{i\bar{i}j\bar{j}}^{a\bar{a}b\bar{b}}$) and sextuples ($D_{i\bar{i}j\bar{j}k\bar{k}}^{a\bar{a}b\bar{b}c\bar{c}}$).}
\label{DociHamil}
\end{figure}

\begin{figure}
\includegraphics[width=0.5\textwidth]{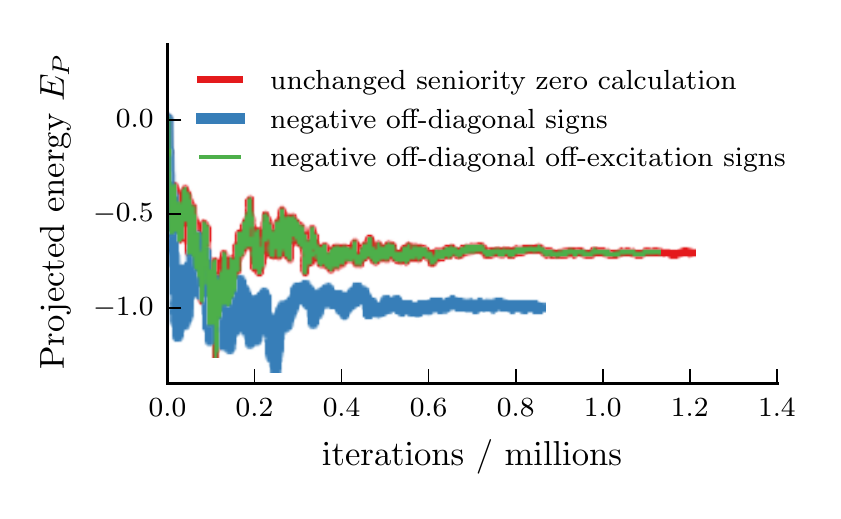}
\caption{A numerical simulation of the 4x5 Hubbard model ($N=10$) to validate the claim of the Hamiltonian structure (Fig. \ref{DociHamil}). 
Here, we treat the system with FCIQMC on just the seniority zero space i.e. DOCI, and the projected energy $E_P$ (Eq.~\ref{proje}) is plotted against the simulation progress in iterations (which is proportionate to imaginary time $\tau$). 
The blue line shows the effect of making all off-diagonal signs negative.
The green line shows a simulation where the signs of the between-excitation-rank blocks (shaded red in Fig. \ref{DociHamil}) are made negative, which then this yields the same diagonalization problem as the original unaltered matrix. The sign convention for the energy is explained in the text.}
\label{DociSignFlips}
\end{figure}

\begin{figure}
\includegraphics[width=0.5\textwidth]{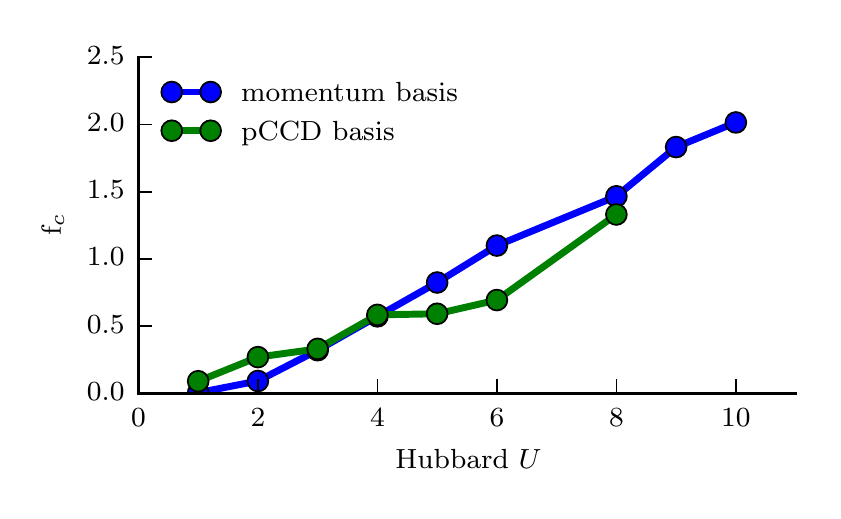}
\caption{The plateau in FCIQMC scales linearly with Hubbard $U$ (for $U>2$), as previously found by Spencer \emph{et al.} (Ref. \onlinecite{spencer_sign_2012}). When a rotation of the orbitals is made to the pCCD basis, the plateau fraction is still comparable for the full FCI space. We show plateau fraction $f_c$, computed as plateau height as a fraction of total size of space. Since the pCCD space is larger due to breaking momentum symmetry, the pCCD basis is more expensive than the original basis to perform FCIQMC. Here, the system size is 4x3, with $N=8$ particles.
}
\label{FciPlateaus}
\end{figure}

\begin{figure*}
\includegraphics[width=0.45\textwidth]{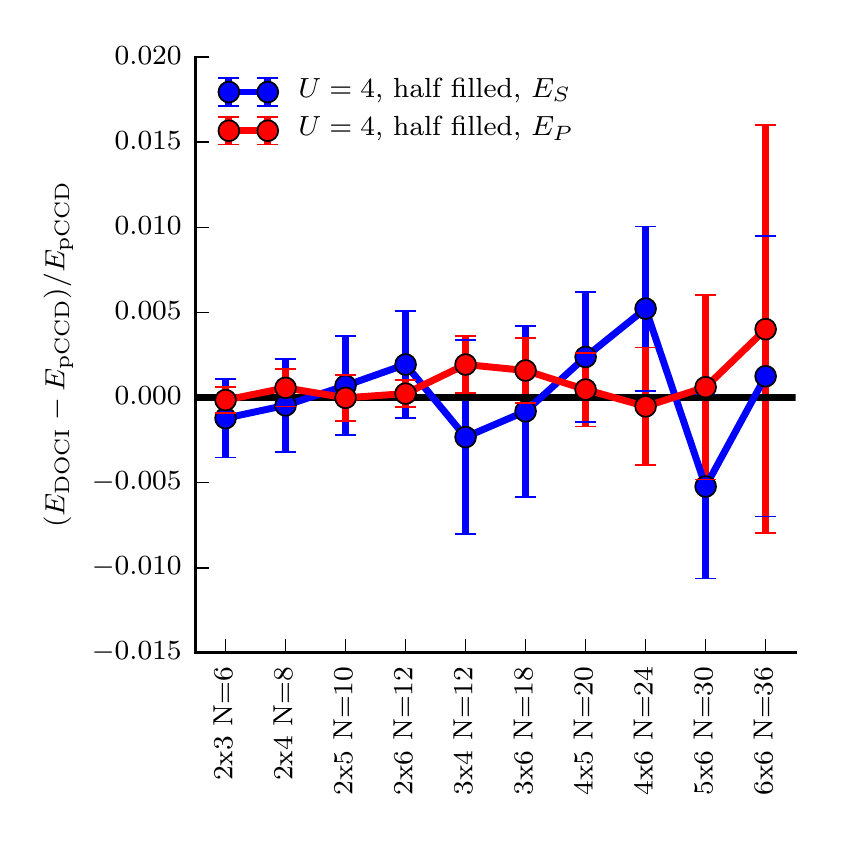}
\hfill
\includegraphics[width=0.45\textwidth]{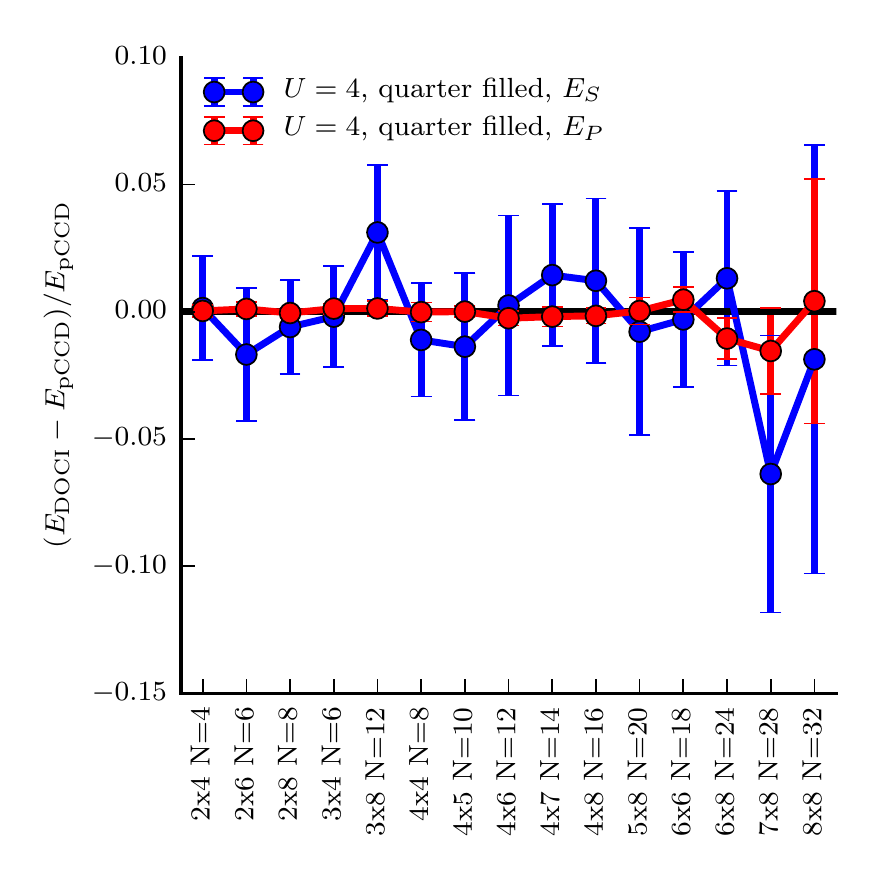}
\caption{There are no statistically significant differences between DOCIQMC energies ($E_\mathrm{DOCI}$) and those from pCCD ($E_\mathrm{pCCD}$) from calculations on a variety of Hubbard models. The two lines refer to two different energy estimators in DOCIQMC: $E_P$ is the projected energy estimator, and $E_S$ is the shift energy estimator.  The left panel shows $N$-particle systems at half filling (filling fraction $f=0.5$), while the right panel shows quarter filling (filling fraction $f=0.25$). Error bars are plotted as $2\sigma$. All energies are plotted rescaled by the pCCD energy.}
\label{PccdCompare}
\end{figure*}

\begin{figure}
\includegraphics[width=0.5\textwidth]{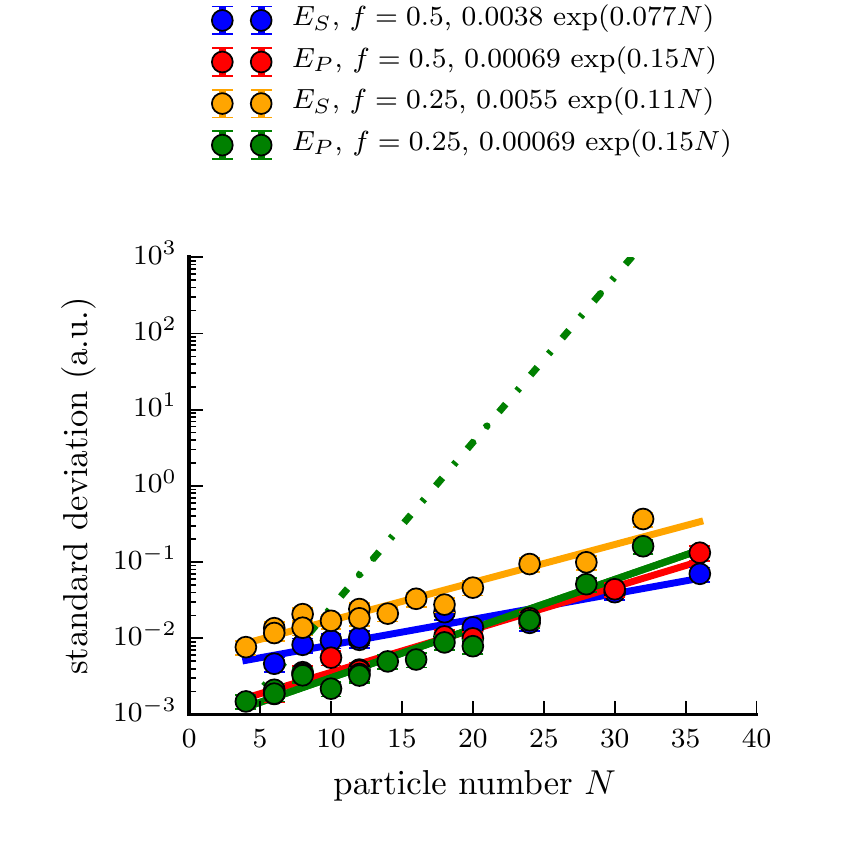}
\caption{The standard deviation for the energy distribution, which determines the stochastic error and cost of the calculation, grows with system size ($N$) as a stretched exponential ($\exp(\alpha N)$). 
Growth of stochastic error scales far slower than if it were to grow with the rising size of the space, and this is shown schematically by the dashed green line. $E_P$ and $E_S$ are the two different measures of the energy; $f$ is the filling fraction. The energies here are absolute, rather than per site.}
\label{errors}
\end{figure}

Restriction of FCIQMC to the seniority zero part of the space has consequences for the sign problem in FCIQMC. 
As mentioned above, the sign problem arises from there being positively and negatively signed off-diagonal Hamiltonian matrix elements, leading to there being both positive and negatively signed walkers in the simulation.
If all off-diagonal matrix elements are negative, the wave function coefficients are positive definite everywhere.
This requires no annihilation to resolve the signs in the wave function and there are no sign problems in this case.

\subsection{The sign problem}

Spencer \emph{et. al}~\cite{spencer_sign_2012} related the FCIQMC sign problem to the energy gap between the real Hamiltonian and a constructed sign-problem-free Hamiltonian where all the off-diagonal signs have been made negative:
\begin{equation}
H^\prime_{ij} = (2\delta_{ij}-1) | H_{ij} | 
\label{FlipSigns}
\end{equation}
whose eigenvector contaminates unconverged simulations (those with fewer walkers than the plateau) in a non-trivial fashion.

Figure \ref{DociHamil} provides a schematic representation of the phenomenon we wish to describe here. 
We can rationalize the existence of such a sign structure by considering that the sign of an off-diagonal matrix element $H_{ij}=\langle D_i |\Psi | D_j\rangle$ derives from the parity of the excitation and the four-index electron repulsion integral. 
This is explained in detail in the paper in which FCIQMC was introduced.\cite{booth_fermion_2009}
The parity is $(-1)^p$, where $p$ is the number of fermion swapping operations required to bring the normal ordered string of $i$ and $j$ into maximal alignment.
When electron pairs are moved, as in the case of DOCI excitations between excitation ranks, these contribute +1 to the parity.
Therefore, one way for the Hamiltonian to have the Fig. \ref{DociHamil} structure is for the matrix elements and parity to all have the same sign for the relevant blocks. 

At first glance, the structure in Fig. \ref{DociHamil} seems to have a severe sign problem. 
Since our propagator is of the form:
\begin{equation}
P=I-\delta \tau H,
\end{equation} 
the presence of positive matrix elements in $H$ leads to a more severe sign problem.

However, consider the following thought experiment. 
Suppose that the double excitation coefficients are all positive. 
The quadruple excitation coefficients would then all be negative, being connected with positive matrix elements to the space of the doubles. The six-fold excitation coefficients are positive, and so on; the structure of the remainder of the wave function follows from this. 

Another way to describe the same simplicity of the sign structure is to consider that all the signs in the red blocks Fig.~\ref{DociHamil} were flipped so that now they are all negative or zero. 
Accordingly, then, the double excitation coefficients are all positive. 
This wavefunction now has the same sign on every determinant -- indicative of a sign-problem-free Hamiltonian --  but the same amplitudes and energy magnitude. 

This structure arises naturally with repulsive two-body interactions in the pairing scheme.  The integrals which connect different excitation ranks are $v_{pq} P_p^\dagger P_q$ and the pair-creation and pair-annihilation operators $P_p^\dagger$ and $P_q$ obey an SU(2) algebra, where
\begin{equation}
P_p^\dagger \, P_q = P_q \, P_p^\dagger + \delta_{pq} \, \left(N_p - 1\right).
\end{equation}
This commutator-based algebra means that no fermionic signs appear, explaining the Hamiltonian structure of Fig. \ref{DociHamil}.
This simple sign structure implies a greatly reduced sign problem and we do not see a substantial sign problem in this space for the system sizes that we examine.

\subsection{Numerical experiment to demonstrate the nature of the sign problem}

Since the foregoing discussion may have been somewhat abstract, we also wish to substantiate what we are saying by means of a numerical simulation. 
Figure \ref{DociSignFlips} shows numerical evidence of our claim that the Hamiltonian has the structure shown schematically in Fig. \ref{DociHamil}. 
Each line represents a single FCIQMC calculation in the seniority zero space started from the same initial conditions and random number seed. 
If no modifications were made, then the lines would be on top of one another.
The red line reflects FCIQMC run as normal on the seniority zero space. 

There are then two further simulations.
In the case of the blue line, all of the off-diagonal signs in the matrix are made negative (Eq. \ref{FlipSigns}).
The converged eigenvalue is more negative than the original eigenvalue. This energy gap is one component of the manifestation of a sign problem in FCIQMC.\cite{spencer_sign_2012}
In the case of the green line, the procedure in Eq. \ref{FlipSigns} is carried out, but only when the matrix elements are between different excitation ranks, \emph{i.e.} flipping the signs of the red off-diagonal blocks in Fig. \ref{DociHamil}.
The correlation energy requires a change in sign, here, to be negative but besides this outcome is an unchanged simulation, and the green line overlaps the red line. 
The implication of this discussion is that the sign problem is substantially weaker in the seniority zero space. 

Despite the sign problem being greatly reduced for the seniority zero space, Fig.~\ref{FciPlateaus} shows that the sign problem is not alleviated beyond the subspace. 
Indeed, the plateau height in walkers as a fraction of the size of the Hilbert space corresponds fairly well to the original fraction required. 
The new space  -- marked on the figure as the pCCD space, to point out that this space is not just truncated but has a rotated basis -- is larger because momentum symmetry has been broken. 
The original basis had a size of space of 22,000 determinants whereas the rotated basis has 250,000.
The change of basis has therefore made the problem harder overall. 
Changes of basis set do in general affect the sign problem in a more non-trivial way in other systems, and this has been studied elsewhere.\cite{schwarz_insights_2015,thomas_symmetry_2014}

\subsection{Discussion}

In this section, we have made an attempt to characterize and describe the sign problem of FCIQMC in the seniority zero space. We have now formulated FCIQMC in the seniority zero space, which we term DOCIQMC for brevity. 

In summary: \emph{if} $\langle pp | v | qq \rangle$ matrix elements are of uniform sign between excitation ranks and derive from a repulsive interaction; \emph{and} the DOCI wavefunction has amplitudes on the double excitations of a uniform sign, \emph{then} the sign problem is greatly alleviated in this space.
Analysis of this is challenging because we cannot assume very much a priori about the structure of the optimized orbitals. 
It is possible to find a Hamiltonian that looks as though it agrees with the above statement, however, there are then two levels of further optimization to move through -- one at the Hartree--Fock level, and then the second at the pCCD level. 
In particular, then, attractive Hamiltonians break from this trend significantly, and are discussed in Sec.~\ref{EnergyDiscussion}.
In this case, therefore, we content ourselves with numerical evidence.

Considering the actual calculations themselves, after the first iteration in solving the pCCD equations, a first order M{\o}ller-Plesset type wavefunction is yielded where the amplitudes are each of opposite sign to the matrix elements \emph{i.e.} $t\propto -v$ if there is a well-defined and unchanging sign for the energy denominator.
If coupling in the quadruples is weak, and the energy denominator does not change sign, then these signs would be consistent with the prior statements in this section. 

This is to say that we might expect the sign problem in DOCIQMC to be weak when the pCCD wavefunction provides a good description of the problem. This seems likely related to the similarity between the structure of the DOCIQMC wavefunction and the pCCD wavefunction, which are, respectively:
$\exp \left( -H \tau \right) | 0 \rangle$ and $\exp \left( T \right) | 0 \rangle$.

\section{Energy comparison between DOCI and pCCD}

Having formulated DOCIQMC, we would now like to use it to study a question of considerable current interest -- does pCCD provide the FCI energy in the seniority zero space?  Prior work has found good agreement generally,\cite{Limacher2013,Henderson2014b,Limacher2014} but also some important discrepancies.\cite{JorgeBreakdown,Henderson2014}  We wish to address this question for system sizes out of reach (due to computational demands) of conventional seniority zero FCI. System size is important because changing system sizes is one means to approach the thermodynamic limit of a problem, which is to say that in reality we want the answer to the question of whether pCCD and DOCI agree for an infinite system.

For a variety of Hubbard systems at $U=4$, we have compared the pCCD energy and the DOCIQMC energy in Fig. \ref{PccdCompare}.  We have selected $U=4$ as a typical case for which strong correlation effects are important, yet simple broken-symmetry mean-field calculations are of insufficient accuracy.\cite{LeBlanc_solutions_2015}  Put differently, $U=4$ represents nearly a worst-case scenario.  We should note here that the coincidence between pCCD and DOCI has been established for systems with repulsive two-body interactions, as we consider here, but seems to break down in systems with attractive two-body interactions.\cite{Henderson2014}  To what extent the poor results of pCCD for attractive systems is relevant for realistic, repulsive Hamiltonians is an open question.

The improved scaling of DOCIQMC means we can approach systems of up to 8x8 for a quarter filling and 6x6 for half filling at modest computational cost. 
Due to the reduced sign problem, the initiator approximation is not used and instead simulations are converged directly.  Having obtained the QMC data, the pCCD energy was subtracted from it and compare this with the size of the error bar.  All but one point lies inside $2\sigma$ from zero (the $\sim$95\% confidence interval), representing reasonable support for the hypothesis that DOCIQMC and pCCD agree.  These results show what we hope to see: that the coincidence between DOCI and pCCD extends to systems of reasonable size, thereby substantiating the idea of pCCD as an affordable and accurate approximation to DOCI across a wide range of systems with a wide range of sizes.

The hinderance of this method is no longer a substantial sign problem, but instead the stochastic error.  The  size scaling of stochastic error in FCIQMC and related algorithms has not been well studied.  To examine this, we would want to look at the standard deviations as a function of system size.  The standard deviation ($s$) is related to the error ($\sigma$) by $s=\sigma N_\mathrm{block}^\frac{1}{2}$, where $N_\mathrm{block}$ is the number of independent data points obtained during a simulation following a re-blocking analysis.\cite{flyvbjerg_error_1989}  This measure is then system-dependent rather than simulation-dependent. 

In Fig. \ref{errors} we show a plot of the standard deviation for the different data series in Fig. \ref{PccdCompare} and find that this grows with system size far slower than the typical exponential growth of the size of the space.  Each data series is fit to a stretched exponential ($\exp(\alpha N)$).  The dashed green line shows the projected growth of the standard deviation if it were to grow in proportion to the size of the space (from $N=4$, for $E_P, f=0.25$).  Therefore, the error grows substantially slower than the $\alpha=1$ conventional exponential growth, and this type of scaling is relatively common in other aspects of FCIQMC and its initiator adaptation.

\subsection{Discussion}\label{EnergyDiscussion}

Here, we have shown the equivalence of the energy for pCCD and DOCI for a wider range of systems -- in particular larger Hubbard models beyond the reach of conventional DOCI.
We should emphasize that pCCD and DOCI are not formally equivalent methods.  For systems with repulsive two-body interactions, pCCD and DOCI deliver very similar wave functions and thus very similar energies, as we have discussed earlier.  Put differently, one observes that with repulsive two-body interactions, the higher-order excitations in DOCI are well-approximated by the exponential of the double excitations.  The same is not true for attractive two-body interactions, where DOCI and pCCD can give very different results~\cite{Dukelsky2003,Henderson2014}.  This is not to say that the higher-order excitations in DOCI cannot be expressed in terms of the double excitations; this can indeed be done, but to do so one must abandon the exponential parameterization of the wave function~\cite{Dukelsky2003,Degroote2015}.

\section{Conclusions}
In conclusion, we formulated a stochastic version of doubly occupied configuration interaction using the FCIQMC algorithm.  In so doing, we have been able to demonstrate that the close resemblance between DOCI and pCCD extends to systems far beyond those for which simple deterministic DOCI can be performed.  While this result is not unexpected, it is encouraging.  The utility of pCCD for the description of strongly correlated systems rests upon three pillars: we require that DOCI with optimized orbitals provides a reasonable description of strongly correlated systems, we require that pCCD give results substantially similar to those of DOCI, and we require that pCCD is able to yield a reasonable set of orbitals in which to describe static correlations.  Our work here confirms that the second of these three criteria remains satisfied for systems far beyond the scope of conventional DOCI calculations.

The sign problem in the seniority zero space is much reduced compared to the full space, but unfortunately this cannot be exploited to alleviate the larger space sign problem.
We gave an explanation for this in terms of the consistently signed blocks in the Hamiltonian and resultant signs of the wavefunction.  The stochastic error is found to be the new bottleneck of the calculations and emerges as a reduced exponential scaling $\exp(\alpha N)$ with system size.

\begin{acknowledgments}
This work was supported by the U.S. Department of Energy, Office of Science, Office of Basic Energy Sciences, Heavy Element Chemistry Program under Award Number DE-FG02-04ER15523. JJS thanks the Royal Commission for the Exhibition of 1851 for a Research Fellowship and computer resources from the Swiss National Supercomputing Centre (CSCS) under project ID s523.  We would also like to thank James S.~Spencer and the developers of the HANDE QMC software package (hande.org.uk).  G.E.S. is a Welch Foundation Chair (C-0036).
\end{acknowledgments}

\end{document}